\documentclass[aps, twocolumn]{revtex4}
\usepackage{graphicx,amsfonts}

\def \be{\begin{equation}}
\def \ee{\end{equation}}
\def \bes{\begin{eqnarray}}
\def \ees{\end{eqnarray}}

\newcommand{\Cb}{{\rm \bf C}}

\def \sl2{SL(2,\Cb)}

\begin{document}

\title{The Feynman propagator for spin foam quantum gravity}

\author{{\bf Daniele Oriti}} 
\affiliation{\small Department of Applied Mathematics and Theoretical Physics, \\
University of Cambridge, Wilberforce Road, Cambridge CB3 0WA, England, EU \\ e.mail: d.oriti@damtp.cam.ac.uk}
\date{\today}

\begin{abstract}
We link the notion causality with the orientation of the 2-complex on which spin foam models are based. We show that all current spin foam models are orientation-independent, pointing out the mathematical structure behind this independence. Using the technology of evolution kernels for quantum fields/particles on Lie groups/homogeneous spaces, we construct a generalised version of spin foam models, introducing an extra proper time variable and prove that different ranges of integration for this variable lead to different classes of spin foam models: the usual ones, interpreted as the quantum gravity analogue of the Hadamard function of QFT or as a covariant  definition of the inner product between quantum gravity states; and a new class of causal models, corresponding to the quantum gravity analogue of the Feynman propagator in QFT, non-trivial function of the orientation data, and implying a notion of \lq\lq timeless ordering\rq\rq.
\end{abstract}
\maketitle

\section{Introduction}
Spin foam models \cite{review,perez} are a new more rigorous implementation of the path integral or sum-over-histories
approach to quantum gravity, in which spacetime is replaced by a combinatorial 2-complex, and the
histories of the gravitational field are characterized by data
taken from the representation theory of the local gauge
group of gravity, i.e. the Lorentz group, and no familiar notions
of metric on differentiable manifolds are used. The main object that
such a sum-over-histories formulation defines is the partition
function of the theory, and from this, allowing the underlying
spacetime 2-complex to have boundaries, quantum
amplitudes functions of the boundary data, that, when a canonical
interpretation is available, should be interpreted as transition
amplitudes between quantum gravity states. The partition function
in these models, for a given 2-complex $\sigma$, has the general
structure: \be
Z(\sigma)=\sum_{\{\rho_f\}}\prod_f\mathcal{A}_f(\rho_f)\prod_e\mathcal{A}_e(\rho_{f\mid
e})\prod_v\mathcal{A}_v(\rho_{f\mid v}), \nonumber \ee where the amplitudes
for faces $f$, edges $e$ and vertices $v$ of the 2-complex are all
functions of the representations $\rho_f$ of the Lorentz group
associated to the faces of the complex.
This work deals, in the spin foam setting, with two related issues common, but not restricted, to any sum-over-histories
formulation of quantum gravity: 1) given a definition of transition amplitudes for
quantum gravity states, which kind of transition amplitude
does it provide? 2) is it possible to identify a notion of
causality and to implement some causality conditions explicitly in
quantum gravity, in spite of the necessary (because
of diffeomorphism invariance) absence of any time parameter in the
quantities one wants to compute? The two issues are related because it is known \cite{halliwell} that in the case of free
relativistic particle and quantum field theory the main difference
between the different transition amplitudes one can construct, in particular the Feynman propagator
and the Hadamard function, is exactly the way they encode (or fail
to encode) causality restrictions, i.e. the fact that one of their arguments 
is in the causal future of the other, or in other words
the {\it order} between these arguments. This order-dependence can be
imposed in each of the histories summed over in a clear way using
a proper time formulation \cite{halliwell,teitelboim}: starting from the same proper time
dependent expression $g(x_1,x_2,T)$, the Hadamard function is obtained
by integrating over both positive and negative
values of the proper time; the resulting amplitude is then a-causal, and
does not register any ordering between its arguments; the Feynman
propagator is instead obtained restricting this integration to
positive proper times only, and this ordering is
precisely what makes it {\it causal}.
What about quantum gravity? Here no external time coordinate is available, so no definition of the
Feynman propagator in terms of time-ordered products is possible;
however, as we said, the crucial property that
distinguishes the two Green functions of the relativistic particle is the symmetry under
a change in the {\it orientation} of the history connecting the
initial and final states and, when a proper time formulation is
available, this is what the signs of the proper time amounts to.
Therefore we see a way out of the mentioned difficulty in the
quantum gravity case. Indeed it has been proposed \cite{teitelboim} that a similar situation as in the
particle case arises in quantum gravity when formulated in the
path integral formalism; the distinction between the
different transition amplitudes is again
obtained by choosing different range of integration in the proper
time formulation of the theory; using a canonical decomposition of
the metric variables, the role of proper time is played by the
lapse function; again an unrestricted range of integration leads
to the (quantum gravity analogue of the) Hadamard function, while
a restriction to positive lapses leads to the (analogue of the)
Feynman propagator \cite{teitelboim}. The problem we address in
this work is how to construct the analogue quantities in the spin
foam context, i.e. in a purely combinatorial, algebraic and group
theoretic way, in absence of any smooth manifold structure and any
metric field. 

\section{Causality as orientation}
In spin foam models spacetime is replaced
by a combinatorial 2-complex, i.e. by a collection of vertices,
links and faces, with extra
data assigning more geometric information to it. What can be the analogue of causal relations in such a
context? Consider
just the first layer of the spin foam 2-complex, i.e. only
vertices and links connecting them, therefore forming a graph. If we add to it orientation data, i.e. arrows on the links,
we obtain an oriented (or directed) graph, a set of oriented links connecting a
set of vertices. Now the vertices can be interpreted as a set of
fundamental spacetime events and the oriented links are then the
causal relations between them. We can assign an {\it orientation
variables} $\alpha_{e\mid v}=\pm 1$ to each link, with respect to each
vertex $v$ it connects, and $\mu=\pm 1$ to each vertex. The spacetime interpretation of these
variables is that of indicating whether the vertex is a future
pointing or past pointing contribution to the overall spacetime
diagram, and whether the link is ingoing or outgoing with respect to each vertex. A consistency condition
for the assignment of orientation data to the graph is that when a
link $e$ connects two vertices it has the opposite orientation in
the two: $\alpha_{e\mid v_1}\mu_{v_1} = - \alpha_{e\mid v_2}\mu_{v_2}$. We also assign an extra
orientation variable to each face $\epsilon_f=\pm 1$. This structure is
basically that of a poset or causal set \cite{Sorkin}, but 
the set of vertices-events endowed with the ordering relation
now fails to satisfy in general any
reflexivity condition, in other words,
our causal relations allow for closed timelike loops. Also, while the combinatorial structure above is general, its causal interpretation makes sense only in a Lorentzian context; therefore
the issue we will be confronting in the following is the general one
of constructing spin foam models that take into
account appropriately the {\it orientation} of the underlying
2-complex, i.e. of {\it orientation-dependent transition
amplitudes} for quantum gravity. In a Lorentzian context
these will have the interpretation of causal amplitudes or of
quantum gravity analogues of the Feynman propagator. The
2-complexes used in spin foam models are not generic: they are
topologically dual to simplicial n-dimensional complexes: to each
vertex corresponds a n-simplex, to each link a (n-1)-dimensional
simplex, to each face an (n-2)-dimensional simplex. The
orientation data we assigned have then a clear geometric
interpretation in this simplicial picture: the $\mu_v$ variable
for a vertex takes the values $\pm 1$ according to whether the
n-simplex dual to it is isomorphic to a n-simplex in Minkowski
(Euclidean) space or the isomorphism holds for the opposite
orientation; the variable $\alpha_{e\mid v} =\pm 1$ indicate
whether the normal to the (n-1)-simplex dual to the link $e$ is
ingoing or outgoing with respect to the n-simplex dual to $v$, and
the variables $\epsilon_f$ also characterize the orientation of
the (n-2)-simplex dual to the face $f$. A second consistency condition on the values of the orientation variables, that basically follows from Stokes's theorem, is then \cite{lo, barrettasym}: 
$\forall v \;\;\epsilon_{f\mid v} = \alpha_{e_1 \mid
v}\alpha_{e_2\mid v}\mu_v$, where $e_1$ and $e_2$ label the two
links that belong to the boundary of the face $f$ and touch the
vertex $v$. This orientation structure is the quantum seed of the classical causal structure, in a Lorentzian context.

\section{Orientation-independence of current spin foam models}
We have now to see how the orientation data introduced in the previous
section enter in current spin foam models. Our analysis holds for the Ponzano-Regge \cite{freidel} and all Barrett-Crane-like models \cite{review,perez} based on ordinary Lie groups and homogeneous
spaces, with no explicit quantum group structure.
Physically these can be interpreted as models of quantum gravity
without cosmological constant. We use a 'first order' formulation of spin foam
models \cite{lo}, in terms of both group variables (or
variables with values in an homogeneous space) and representation
variables. 
Let us anticipate the result before justifying it: all current spin foam models do not depend, in their amplitudes, on the orientation of the underlying 2-complex. 
The way this is achieved is quite simple: in the expression for the amplitudes for spin foams the terms that can be understood as contributions from opposite orientations are summed simmetrically thus erasing the dependence on the orientation data. We deal here explicitely only with the n-dimensional models for $n\geq 4$ based on the homogeneous spaces $SO(n-1,1)/SO(n-1)\simeq H^{n-1}$ and $SO(n)/SO(n-1)\simeq S^{n-1}$ for brevity, but the analysis can be extended with the same results for all the other spin foam models \cite{oriented}. 
The Barrett-Crane-like models in 4 and higher dimensions \cite{review} take the form:
\bes
Z=\left(\prod_f \int d\rho_f\right)\left(\prod_v \prod_{e\in v} \int_{H_e}dx_e \right)\prod_f \mathcal{A}_f(\rho_f) \nonumber \\ \prod_e\mathcal{A}_e(\rho_{f\in e})\prod_v\mathcal{A}_v(x_{e\in v},\rho_{f\in v}) \nonumber
\ees
where of course the precise combinatorics varies according to the dimension, but in any case: $\rho_f$ are the unitary irreps of the local gauge group of gravity ($SO(n-1,1)$ or $SO(n)$), and these (class I) unitary representations are labelled by either a half-integer $j$ in the Riemannian case, or by a positive real parameter $\rho$ in the Lorentzian, $H_e$ is the homogeneous space to which the vectors $x_e$, interpreted as normals to (n-1)-simplices in each n-simplex, belong.  
The vertex amplitudes factorize in terms of face contributions, and it is at this level that the information about the orientation of the 2-complex is erased:
\bes
\lefteqn{\mathcal{A}_v(x_{e\in v},\rho_{f\in v})=\prod_{f\in v}\mathcal{A}_{f\in v}(\theta_f,\rho_f)=} \nonumber  \\ &=&\prod_{f\in v}\left( \mathcal{W}_f^{\mu_v=+1}(\vartheta_f,\rho_f) + \mathcal{W}_f^{\mu_v=-1}(\vartheta_f,\rho_f) \right) \nonumber
\ees
where the amplitudes depend on the $x_e$ vectors only through the invariant distances $\theta_f=\cosh(h)^{-1}(x_{e1\in f}\cdot x_{e2\in f})$, and in the last step we have traded the orientation data on the faces for those on the vertices by using the second consistency relation, indicating as $\vartheta_f$ the oriented angle $\alpha_{e1}\alpha_{e2}\theta_f$. 
This orientation independent structure is pretty general (independent on signature or dimension), and such are the mathematics behind this and the orientation {\emph dependent} functions $\mathcal{W}$.
In fact the amplitudes $\mathcal{A}_{f\in v}$ are constructed out of representation functions of the relevant Lie group, $\mathcal{A}_{f\in v}=D^{\rho_f}_{00}(\theta_f)$, i.e. they are given by zonal spherical functions. In turn all these representation functions can be \emph{uniquely} expressed in terms of sums of so-called \lq\lq representation functions of the 2nd kind\rq\rq $E^{\rho_f}_{kl}(g)$ \cite{ruhl}, solutions of the same set of equations \cite{V-K} with different boundary conditions \cite{V-K}, and it is precisely these representation functions of the 2nd kind that originate the orientation-dependent amplitudes $\mathcal{W}_{f\in v}$ \cite{oriented}. The relation between representation functions of the 1st and 2nd kind is best understood in terms of their analytic expressions in terms of associated Legendre functions of the 1st and 2nd kind \cite{V-K, oriented}:   
\bes
\lefteqn{\mathcal{A}_{f \in v}=D^{\rho}_{00}(\theta_f)= \frac{2^{\frac{n-3}{2}}\Gamma\left(\frac{n-1}{2}\right)}{(\sin(h)\theta)^{\frac{n-3}{2}}} {\Large P}^{\frac{3-n}{2}}_{\sigma}(\theta)=} \nonumber \\ &=& \frac{2^{\frac{n-3}{2}}\Gamma\left(\frac{n-1}{2}\right)}{(\sin(h)\theta)^{\frac{n-3}{2}}}\frac{e^{-i\pi\frac{n-3}{2}}}{\Gamma\left( \sigma + \frac{n-1}{2}\right)\Gamma\left( -\sigma -\frac{n-3}{2}\right)\cos\pi\sigma}\times \nonumber \\ &\times& \left[ {\Large Q}^{\frac{n-3}{2}}_{-\sigma -1}(\theta)\,-\,{\Large Q}^{\frac{n-3}{2}}_{\sigma}(\theta)\right] \,=\, E^{\rho}_{00}(\theta)\,+\,E^{-\rho}_{00}(\theta)\,= \nonumber \\ &=&\mathcal{W}_f^{\mu_v=+1}(\vartheta_f,\rho_f) + \mathcal{W}_f^{\mu_v=-1}(\vartheta_f,\rho_f) \label{splitting} \nonumber
\ees
where the $\sigma = 2j + (n-3)/2$ in the n-dimensional Riemannian case, and $\sigma = i\rho - 1/2$ in the Lorentzian case.    
A similar relation can be obtained \cite{oriented} for the 3-dimensional Ponzano-Regge case, and for the other Lorentzian models. 
This orientation independence leads to interpreting the current spin foam models as a-causal transition amplitudes, i.e. as the quantum gravity analogue of the Hadamard function; when a canonical formulation is available, they can be equivalently thought of as defining the physical inner product between quantum gravity states, invariant under spacetime diffeomorphisms, or as a covariant definition of the matrix elements of the projector operator onto physical states \cite{carloprojector}. 
Given the universal structure outlined above, we are lead to look for a universal way of modifying current spin foam models and to a new, again universal, definition of orientation-dependent or causal spin foam models \cite{oriented}; these would correspond to causal transition amplitudes and to a quantum gravity analogue of the Feynman propagator of field theory. The brute force procedure is obvious and was performed in the 4-dimensional Lorentzian case in \cite{lo}: simply restrict the spin foam amplitudes switching from D functions to E functions corresponding to a consistent choice of values for the orientation data, dropping the sum over $\mu_v$ or $\epsilon_f$. Hovewer, we will shortly see that a much more natural and elegant construction exists. Also, the expressions \ref{splitting} are reminescent of the decomposition of Hadamard functions into Wightman functions. The new construction we are proposing shows that this analogy is indeed exact.

\section{Spin network feynmanology and particles on Lie groups/homogeneous spaces}
The amplitudes assigned to spin foam faces, edges and vertices in the Barrett-Crane-type models is given by the evaluation of simple spin networks and was described in \cite{FK} in analogy to the evaluation of Feynman diagrams: assign a variable valued in the relevant homogeneous space to each vertex of the given spin network, assign a zonal spherical function $D^{\rho_f}_{00}(\theta_f)$  to each line and sum over all the possible values of the variables on the vertices to get the final amplitude. In this prescription the zonal spherical function is treated as a kind of propagator, and this is what it really is, as it turns out.
Consider a scalar field $\phi(g)$ with mass $m$ living on the Lie group/homogeneous space $G$, with each point on it labelled by $g$; consider its free evolution parametrised by a proper time coordinate $s$; the equation of motion in proper time is: $(i \partial_s + \Delta) \phi(g,s) = 0$ with $\Delta$ being the Laplace-Beltrami operator on $G$. The dynamics is completely captured by the evolution (heat) kernel $K(g,g',s)=K(g g_{-1},s)$ \cite{camporesi,marinov}, in the sense that given the initial condition $\psi(g_0,0)$, we have: $\psi(g,s)=\int dg_0 K(g,g_0,s)\psi(g_0,0)$. The various physical propagators are obtained from the evolution kernel according to how the proper time is integrated out; the Hadamard function is obtained via the expression: $H(g,g',m^2)= -i \int_{-\infty}^{+\infty}ds K(g,g',s)e^{-i m^2 s}$, while restricting the range of integration to positive proper times only gives the Feynman propagator $G_F(g,g',m^2)= -i \int_0^{+\infty} ds K(g,g',s) e^{-i m^2 s}$, where the usual Feynman prescription $m^2 \rightarrow m^2 -i\epsilon$ is assumed for convergence.    
It turns out \cite{oriented} that the functions entering the expressions for the quantum amplitudes of all current spin foam models correspond to the Hadamard 2-point functions for a scalar field on the relevant Lie group/homogeneous space with $m^2=-C(\rho_f)$ where $C(\rho_f)$ is the Casimir eigenvalue of the representation labelling the face of the 2-complex (therefore the link of the spin network whose evaluation gives the amplitude for the vertex):
\bes   
\lefteqn{H(\theta_f, m^2)=\frac{i}{2\pi}\sqrt{\Delta_{\rho_{f\in v}}} D^{\rho_{f\in v}}_{00}(\theta_{f})=} \nonumber \\ &=&-i\int_{-\infty}^{+\infty}ds_{f\mid v} K_{\mathcal{M}}(\vartheta_f,T_v)e^{+iC(\rho_f)T^f_v}, \nonumber \ees 
with the \lq\lq oriented proper time\rq\rq being $T=\mu_v s$, $\Delta_{\rho}$ is the dimension of the representation $\rho$ in the Riemannian case, or in the Lorentzian case (where the unitary representations are infinite dimensional) the contribution of the representation to the Plancherel measure, and the Casimir eigenvalues are $C(\rho)=2j(2j+n-2)$ with $j$ half-integer in the Riemannian case, and $C(\rho)= +\rho^2 +(n-2/2)^2$ with $\rho$ positive real in the Lorentzian case based on the timelike hyperboloid. Again a similar formula holds for the Ponzano-Regge models and for the models based on the (n-1)-dimensional DeSitter space. It is clear that the result is independent of the value of the various orientation data.
On the other hand simply imposing the above restriction in the proper time integration (that amount indeed to a causality restriction for the particle evolution \cite{teitelboim}), one obtains the expression for the Feynman propagator that one needs to define orientation dependent or causal spin foam models, and its expression in terms of representation functions of the 2nd kind, confirms their interpretation as Wightman functions:
\bes
\lefteqn{G(\vartheta_f, m^2,\mu_v)=-i\int_{0}^{+\infty}ds_{f\mid v} K_{\mathcal{M}}(\vartheta_f,T_{f\mid v})e^{+iC(\rho_f)T_{f\mid v}}=} \nonumber \\ &=\frac{i}{2\pi}\sqrt{\Delta_{\rho_{f}}}\left[ \theta(\mu_v)E^{\rho_{f}}_{00}(\vartheta_{f})+\theta(-\mu_v)E^{-\rho_{f}}_{00}(\vartheta_{f}) \right]= \nonumber \\ &=\frac{i}{2\pi}\sqrt{\Delta_{\rho_{f}}}\left[ \theta(\mu_v)\mathcal{W}^{+}(\vartheta_{f},\rho_f)+\theta(-\mu_v)\mathcal{W}^{-}(\vartheta_{f},\rho_f) \right]\,\,\,\,\nonumber.
\ees Similarly for the other spin foam models \cite{oriented}.
This is a non-trivial function of the orientation data, with the usual \lq\lq time ordering\rq\rq  being replaced by a \lq\lq timeless ordering\rq\rq !

\section{A proper time formalism and the Feynman propagator for spin foam quantum gravity}
Taking seriously this particle analogy and the associated spin network feynmanology, and having also in mind the group field theory approach \cite{review}, we can generalise the current formulation of spin foam models to include a proper time variable, and obtaining a general expression from which both orientation independent and causal models can be derived, simply changing the integration contour in proper time. The expression for the models dealt with above looks as follows:
\bes 
\lefteqn{Z= \prod_v \prod_{f\mid v}\int_C ds_{f\mid v} \prod_f \sum_{\rho_f} \prod_v \prod_{e \mid v} \int_{H_e} dx_e \, {\large \mathcal{A}(\rho_f, x_e, T_{f\mid v})}}\;\;\;\;\;\;\; \nonumber \\ &{\large \mathcal{A}(\rho_f, x_e, T_{f\mid v})}= \prod_f \mathcal{A}_f(\rho_f) \prod_e \mathcal{A}_e(\rho_{f\mid e})\;\;\;\;\;\;\;\;\; \nonumber \\ &\prod_v\prod_{f\mid v} \left( -2\pi i \left(\sqrt{\Delta_{\rho_f}}\right)^{-1} \,\,K(\vartheta_{f\mid v},T_{f\mid v}) e^{i C(\rho_f) T_{f\mid v}}\right)\;\nonumber.    
\ees 
A similar formula holds for all the other models.
This generalised expression encompasses both the usual un-oriented models and the new causal ones; indeed the first are obtained by choosing the extended range of integration $C=(-\infty, +\infty)$ for the $T$ variable (and this erases the dependence on $\mu_v$ and $\alpha_e$, while the quantum gravity Feynman propagator is obtained with $C=(0,+\infty)$. In this last case, a regularization prescription for convergence is implicit for each variable (so that the expression has to be understood in the complex domain): $\rho_f \rightarrow \rho_f +i \epsilon$, $\vartheta_f \rightarrow \vartheta_f + i \delta$. The explicit form of the evolution kernel $K$ differs of course in the different models and affects the exact form of the amplitudes \cite{oriented}, that however all share the general structure here presented\footnotemark. \footnotetext{The technology related to quantum particles on homogeneous spaces could have been used to generalise and then modify also the face and edge amplitudes. The reason why we have not done so is twofold: on the one hand the usual interpretation of these contributions to the spin foam models is that of a contribution to the overall measure, therefore the implementation of causality may be needed only at the vertex level, that is instead supposed to encode the dynamics of the theory; on the other hand, the form of the edge amplitudes in some version of the 4-d Barrett-Crane model is understood as arising directly from the form of boundary spin network states \cite{boundary}, and we prefer, at this stage, to keep that structure without modification. We believe, however, that alternative formulations of the models and possible definitions of modified spin network states, maybe in order to induce an orientatation-dependence in their structure, deserve further analysis.}    
Let us discuss some properties of these new kind of transition amplitudes. First, all these models can be recast in the form of quantum causal histories models \cite{fotini,lo}; therefore they define quantum amplitudes for causal sets, if the underlying oriented graph does not contain closed timelike loops, and with the additional restriction of the vertices being $(n+1)$-valent in n dimensions. 
Second, while the un-oriented models can be related to the classical Regge action only in a asymptotic limit, when their vertex amplitudes result in being proportional to the cosine of it \cite{barrettasym,fl,b}, here the connection with the Regge action is manifest without any approximation; for example, in the 4-dimensional case the relevant evolution kernel has the form \cite{camporesi}: 
$K(\vartheta,T) = \frac{1}{(4\pi i T)^{3/2}}\left( \frac{\vartheta}{\sinh\vartheta}\right) e^{i\frac{\vartheta^2}{4T} -iT}$ and  
the causal vertex amplitude takes the form:
\bes
\lefteqn{\mathcal{A}^C_v = \prod_{f \in v} \left( -2\pi i \rho_f \int_0^{+\infty} ds_{f\mid v} K(\vartheta_f,T_{f\mid v}) e^{i C(\rho_f)T_{f\mid v}}\right) =\;\;\;\;} \nonumber \\ &=& - \left( \prod_f \frac{1}{\rho_f\sinh\vartheta_f}\right) \,e^{i\sum_{f\in v} \mu_v \rho_f \vartheta_f}\;\;\;\;\;\;\;\;\;\;\;  \nonumber
\ees  
and therefore the product over vertex amplitudes in the spin foam model gives
\bes
\prod_v\mathcal{A}^C_v=\left(\prod_v\prod_{f\in v}\frac{1}{\rho_f\sinh\vartheta_f}\right)\,e^{i\sum_{f\in v} \rho_f \sum_v \mu_v  \vartheta_{f\in v}}= \nonumber \\ = \left(\prod_v\prod_{f\in v}\frac{1}{\rho_f\sinh\vartheta_f}\right)\,e^{i {\Large S_{R}}(\rho_f,\vartheta_f)} \nonumber,
\ees
i.e. the exponential of the Regge action (in first order formalism \cite{lo}) for simplicial gravity, with an additional contribution to the overall measure, as one would expect from a sum-over-histories formulation of quantum gravity based on a simplicial discretization. The same is easily shown in the 4d Riemannian case and in the 3-dimensional case, while the proof for the higher dimensional models is less straightforward \cite{oriented}. Third, the causal/oriented models seem to solve the problem, present in all BF-type formulations of quantum gravity \cite{freidelkrasnov, dp-f}, of different isomorphic sectors related by a change of orientation all summed over in the path integral quantization and thus interfering, leading to a discrepancy with the straightforward quantization of GR, and affecting the computation of geometric quantitites in spin foam models \cite{freidelkrasnov}; in the causal models it seems instead that a restriction to the GR sector is achieved. 

\section{Conclusions}
Let us summarise our results: 1) we have linked the notion of causality in a Lorentzian context with the orientation of the 2-complex on which spin foam models are based, and have identified the relevant data characterizing this orientation; 2) we have shown that all current spin foam models are orientation-independent; 3) using the technology of evolution kernels for quantum fields/particles on Lie groups/homogeneous spaces, we have constructed a generalised version of spin foam models, introducing an extra variable with the interpretation of proper time; 4) we have proven that different ranges of integration for this proper time variable lead to different classes of spin foam models: one corresponds to the usual ones, interpreted as the quantum gravity analogue of the Hadamard function of QFT or, equivalenty, in a canonical interpretation, as a covariant  definition of the inner product between quantum gravity states; the other is a new class of models and corresponds to the quantum gravity analogue of the Feynman propagator in QFT, i.e. a causal transition amplitude, non-trivial function of the orientation data, that implies a notion of \lq\lq timeless ordering\rq\rq; 5) we have shown how the causal model is manifestly related to simplicial gravity in the 4d Lorentzian case. All these results hold true in full generality, for the type of spin foam models considered, i.e. regardless of the spacetime dimension and signature.  
These results open quite a few lines of possible further research. We list some of them: 1) a group field theory formulation of generalised and oriented models (work on this is in progress); 2) an investigation of the possibility of defining a notion of \lq\lq positive and negative energy\rq\rq sectors, also at the level of spin network states, based on orientation/causal properties, in a timeless framework; 3) the study of the role and significance of the proper time formalism, and of the possible link between this  group theoretic proper time parameter and the lapse function of canonical quantum gravity; 4) an analysis of the relationship between the spin foam approach and other approaches to quantum gravity, starting from the new causal models, that seem to be the at the point of convergence of simplicial quantum gravity \cite{williams}, dynamical triangulations (when a sum over 2-complexes is implemented) \cite{loll} and causal sets, in addition to canonical loop quantum gravity \cite{carlobook}; 5) the new models can be interpreted as defining the matrix elements of an \lq\lq evolution operator\rq\rq, the basis for any \lq\lq scattering calculation\rq\rq, whose property could be studied, also in order to understand whether a notion of unitary evolution is feasible in Quantum Gravity, in absence of an external time coordinate, but with a clear notion of causality.

\end{document}